\documentclass[aps,prb,reprint,groupedaddress]{revtex4-2}

\usepackage{amsmath}
\usepackage{amssymb}
\usepackage{mathtools}
\usepackage[english]{babel}
\usepackage{color}
\usepackage{graphicx}
\usepackage{dcolumn}
\usepackage{bm}
\usepackage{hyperref}

\begin{document}

\title{Hysteresis and jumps in the $I-V$ curves of disordered two-dimensional materials}
\author{Shahar Kasirer}
\email[]{kasirer@mail.tau.ac.il}
\altaffiliation{School of Neurobiology, Biochemistry and Biophysics, George S. Wise Faculty of Life Sciences \& Raymond and Beverly Sackler School of Physics and Astronomy, Faculty of Exact Sciences, Tel Aviv University, Tel-Aviv 6997801, Israel}

\author{Yigal Meir}
\email[]{ymeir@bgu.ac.il}
\affiliation{Department of Physics, Faculty of Natural Sciences, Ben Gurion University of the Negev, Beer Sheva 84105, Israel}
\date{\today}

\begin{abstract}
The superconductor-insulator transition (SIT) in thin-film disordered superconductors is a hallmark example of a quantum phase transition. Despite being observed more than 30 years ago, its nature is still under a lively debate. One intriguing observations concerns the insulating side of the transition, which exhibits some unusual properties. Among them is its current-voltage relation ($I-V$ curve), which includes (i) a conductance that changes abruptly by several orders of magnitude with increasing voltage, (ii) hysteretic behavior, and (iii) multiple (sometimes more than 100) smaller current jumps near the transition. Some models have been suggested before, but no model has been successful in accounting for the observed behavior in full. One commonly used approach is to model the disordered sample as a two-dimensional array of conducting islands, where charge carriers tunnel from one island to its neighbors. In those models, fast relaxation is assumed, and the system is treated as always being in electrostatic and thermal equilibrium. Those models are successful in explaining some measurement results, including the phase transition itself, but they fail to reproduce hysteresis in their predicted $I-V$ curves. Here, we suggest incorporating finite relaxation time into an array model. We show that, in the slow relaxation limit, our model can reproduce hysteresis and multiple jumps in the $I-V$ curve. Based on our results, we argue that a similar behavior should also be observed in two-dimensional normal (nonsuperconducting) arrays. This claim is supported by past observations. We analyze the role of different parameters in our model, determine the range of relevant timescales in the problem, and compare our results with selected measurements. 

\end{abstract}


\maketitle

\section{Introduction \label{introduction}}

The superconductor  - insulator transition (SIT) is a hallmark example of a quantum phase transition. In this transition, a superconducting sample is driven into an insulating phase by changing its thickness, disorder, or applying an external magnetic field. This transition is not unique to one specific material, it had been observed in different disordered superconductors, including granular superconductors \cite{frydman2002}, amorphous In:O \cite{hebard1990, sambandamurthy2004} and TiN films \cite{baturina2007-1}. Despite being observed more than 30 years ago \cite{haviland1989,hebard1990}, its nature is still under a lively debate, due to the interplay between superconductivity and disorder \cite{goldman2010}. 

One of the first models proposed for this transition is of a granular superconductor (SC), composed from SC islands \cite{efetov1980}. Cooper-pairs and quasi-particles, which are localized on those islands, can tunnel between neighboring ones. This is essentially a disalso ordered Josephson-Junction array (JJA) model, and had been used in a few theoretical works concerning the SIT \cite{vinokur2008, bard2017, basko2020}. The SIT had been observed experimentally in an actual JJA \cite{delsing1998, bottcher2018}. Previous theoretical works suggest that amorphous, uniformly disorder systems, break into ``islands" with higher SC order parameter, separated by ``insulating" areas where the SC order is weak \cite{ghosal1998,ghosal2001,dubi2007}. Such spatial fluctuations in the SC gap have indeed been observed experimentally \cite{sacepe2008, tan2008, sacepe2011,chand2012, baturina2013, kremen2018}. In addition, various experiments indicate the persistence of SC correlations into the insulating side of the transition \cite{barber1994,kowal1994,crane2007, sacepe2008,tan2008,nguyen2009,kamlapure2010}, further supporting this model.

\begin{figure*}[ht]
    \includegraphics[width=0.72\textwidth]{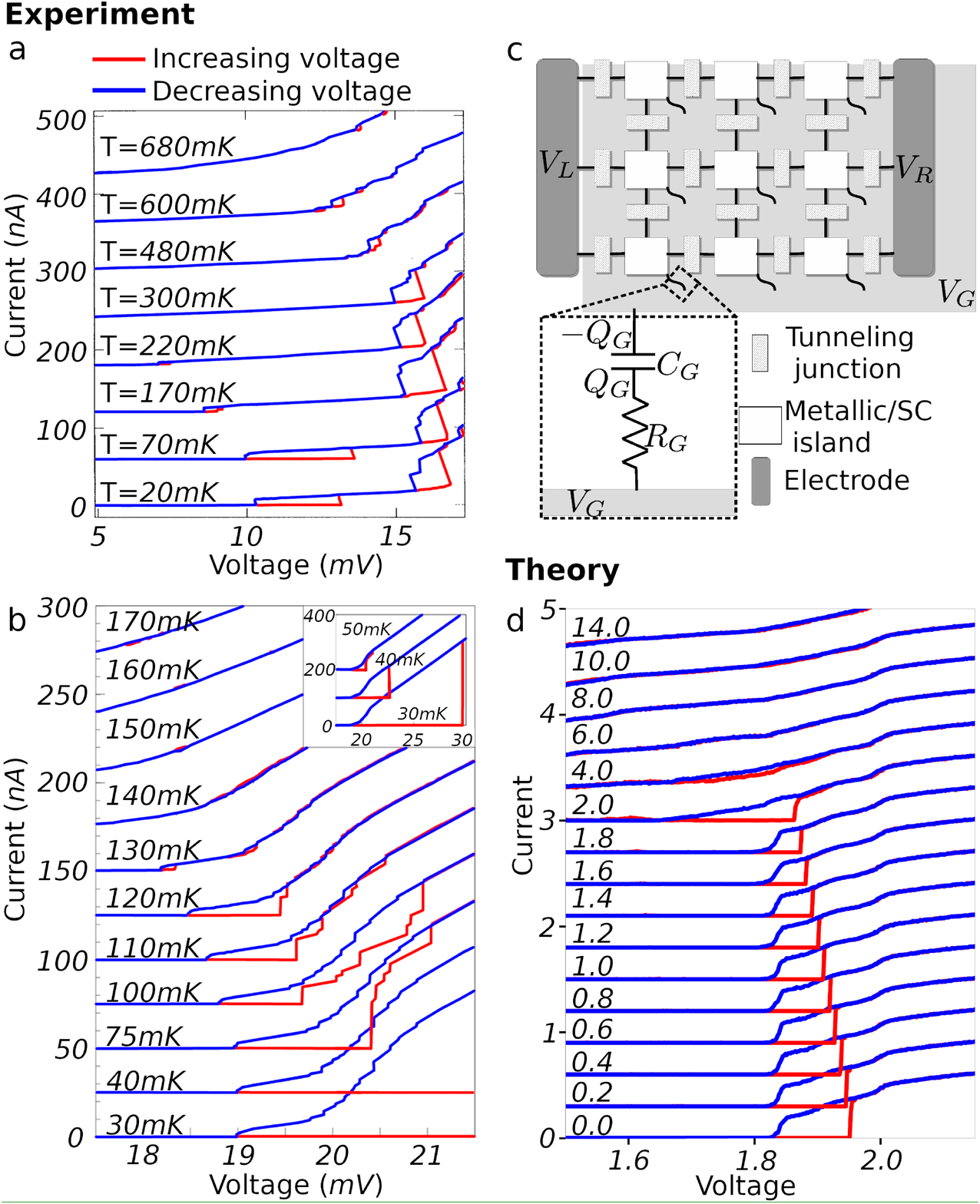}
    \caption[A model of tunneling junctions 2D array can reproduce hysteresis and jumps in its I-V curve]{
    (a) Experimental $I-V$ curves of a $200\times200$ GaAs quantum dot array \cite{duruoz1995}. The voltage is measured directly across the array in a four-lead configuration.
    (b) Experimental $I-V$ curves of an a:InO sample in the insulating phase, near the superconductor-insulator transition, for different temperatures. Curves are offset from one another by 25$nA$. Inset: Expanded $V$-range for the low temperature results \cite{cohen2011}.
    (c) The 2 dimensional island array model: charge carriers can tunnel from one island to its neighbors through the tunneling junctions. Each tunneling junction is attributed a tunneling-resistance and capacitance. Each island is coupled to the gate of voltage $V_G$, via an RC circuit (the inset shows an enlarged view of this coupling).
    Simulated $I-V$ curves of a $10\times 10$ array model for different temperatures. Temperature, voltage, and current are expressed in units of $e^2/\left(k_B\left<C\right>\right)\cdot 10^{-3}$, $e/\left<C\right>$ and $e/\left(\left<R\right>\left<C\right>\right)$, respectively. $\left<C\right>$ and $\left<R\right>$ are the average tunneling junctions capacitance and resistance, respectively. Curves are offset from one another by $0.3e/\left(\left<R\right>\left<C\right>\right)$.
    Simulation parameters are  $V_{max}=2.4\left<C\right>/e$, $V_G= 0$, $R_G=50\left<R\right>$, $C_G = 5\left<C\right>$, $\sigma_C = 0.05\left<C\right>$, $\sigma_R = 0.9\left<R\right>$.
    }
    \label{fig:2D_model}
\end{figure*}

An intriguing observation related to SIT concerns the nonlinear $I-V$ characteristics measured in the insulating regime near the transition. An example of those intriguing experimental observations is depicted in Fig.~\ref{fig:2D_model}b: the  current  is zero, to experimental accuracy, up to some threshold voltage, and then it rises abruptly. This behavior is, in fact, hysteretic, and the threshold voltage depends on the direction of the voltage tuning. Such data have been obtained both in In:O \cite{sambandamurthy2004,ovadia2009, cohen2011} and TiN \cite{baturina2007-1,vinokur2008} films, though the interpretation of the data was quite different. Even more surprisingly, as can be seen in the same figure, the big jump in the current consists, close to the SIT, of multiple (sometimes more than 100 \cite{cohen2011}) smaller current jumps. Multiple jumps have also been observed in a JJA \cite{delsing1994}, a system that also exhibits hysteresis (see e.g. \cite{bottcher2018}). This has led to a further debate concerning these observations, including the suggestion that this is due to overheating of electrons \cite{altshuler2009,ovadia2009,levinson2016}
or the formation of a novel "super-insulator" phase, which is also based on a model of SC islands  \cite{vinokur2008,baturina2013}. Interestingly, similar data were obtained in a quantum-dot array (QDA) \cite{duruoz1995,staley2014} (Fig.~\ref{fig:2D_model}a),  $Y_xSi_{1-x}$ thin films \cite{ladieu1996} and Au nanocluster films \cite{ancona2001}, all non SC materials, which suggests a mechanism not limited to superconductors.  

Using the SC islands model, theoretical $I-V$ relations can be calculated. In such a model one would expect a threshold voltage due to Coulomb blockade, and even multiple current jumps, due to the opening of new parallel transport channels, or allowing multiple occupation states of the islands. Indeed,
previous calculations of the current-voltage characteristics in normal arrays  gave rise to a threshold voltage \cite{middleton1993,enomoto1999,kaplan2003} and even multiple jumps \cite{muller1998,jha2005}. Similar threshold behavior in the insulating phase has also been observed in simulations of superconducting JJAs \cite{otterlo1992,reinel1994}. However, none of these simulations exhibited the hysteresis observed in the experiment.

As mentioned above, another theoretical model that was suggested to explain the measured $I-V$ curves is the ``electron overheating" model \cite{altshuler2009}. In this model, when the applied voltage crosses a threshold value, the electron temperature becomes higher than that of the phonons, manifested by a big jump in the $I-V$ curve. Upon decreasing the voltage, electrons stay overheated down to a second threshold voltage, lower than the first one, leading to hysteresis in the observed curve. Although this model, which uses as input the measured relation between resistance and temperature, $R\left(T\right)$, was successful in reproducing the abrupt jump and the hysteresis in the $I-V$ curve, it could not account for multiple current jumps (at least not with the measured $R\left(T\right)$  relations \cite{levinson2016}). 

Accordingly, none of the existing theories can explain the combined experimental observations of threshold, hysteresis  and  multiple jumps in the $I-V$ curve in the same sample, adding to the enigma of the nature of the insulating state close to the SIT. Here, we build upon the model of a disordered SC islands, and demonstrate how, by incorporating the physically relevant relaxation time, one can explain these observations within a single model. 

\section{Suggested Model \label{model}}

When one considers transport of electrons or Cooper pairs through the array, there are three important times scales: (i) the time it takes for charge carriers to tunnel between islands, (ii) the intertunneling time, i.e., the time between tunneling events, and (iii) the relaxation time, i.e., how fast an array restores electrostatic and thermal equilibrium after a tunneling event has occurred.
In most existing array models, both tunneling and relaxation time scales are assumed to be much shorter than the intertunneling time, and thus they are treated as instantaneous. While this assumption is justified for the tunneling time, which is usually the fastest time scale in the system,
the relaxation time depends on various sample parameters, such as capacitance and resistance to the gates or the substrate. If it is much longer than the intertunneling time, one cannot assume that the system is always in electrostatic equilibrium.  Below we show that in this ``slow relaxation limit", one can reproduce the main features in the experimental $I-V$ curve.

Since the physics we describe also applies to normal arrays, we first start with this simpler 
system \cite{bakhvalov1991,middleton1993,enomoto1999}. A schematic drawing of our model is depicted in Fig. \ref{fig:2D_model}c - a square array of metallic islands, where nearest neighbors are coupled electrostatically and by tunneling, and each island is also coupled to a gate/substrate. We first introduce the calculation of the rates in which electrons tunnel between neighboring islands and then explain how we incorporated a finite relaxation time into the model. To calculate tunneling rates, we use standard perturbation theory. The electrostatic energy of our two-dimensional islands array is given by
\begin{eqnarray}
&E = \sum\limits_{\substack{i \in \text{islands}}}\left[ \frac{C_G}{2}\left(V_i - V_G\right)^2 + \sum\limits_{\substack{j \in \text{neighbors of} \\ \text{island $i$}}} \frac{C_{ij}}{4} \left(V_i - V_j\right)^2\right]  \nonumber \\ & +\sum\limits_{\substack{x \in \text{electrodes}}}\left[Q_x V_x + \sum\limits_{\substack{j \in \text{neighbors of}\\ \text{\hspace*{0.2cm}electrode $x$}}} \frac{C_{xj}}{2} \left(V_x - V_j\right)^2
\right]
\end{eqnarray}
where $V_i, V_x, V_G$ are the electric potentials on island $i$, electrode $x$ (left or right), and the gate, respectively. $Q_x$ is the charge on electrode $x$, and $C_{ij}$ is the capacitance of the tunneling junction connecting islands $i$ and $j$. The factor of $1/4$ in front of the second term is there to avoid double counting. For our calculation, we would be interested in the change in energy resulting from a single tunneling event. For an electron tunneling from island $i$ to island $j$, the change in electrostatic energy can be presented in a simple form \cite{bakhvalov1991}
\begin{equation}
\Delta E^{i\rightarrow j} = \frac{e}{2}\left[\left(V_{j}-V_{i}\right) + \left(V'_{j}-V'_{i}\right)\right]
\label{eq:energy_difference}
\end{equation}
where $V_i$ and $V_i'$ are the electric potentials on island $i$ before and after tunneling, respectively.
To relate the electric potential on each island to the charge on it, we define the capacitance matrix $\mathcal{C}$:
\begin{equation}
\mathcal{C}_{i,j} =
\begin{cases}
\sum\limits_{\substack{k \in \text{neighbors of} \\ \text{island $i$}}} C_{i,k} & \text{if } i=j\\
-C_{ij} & \text{if island $i$ and $j$ are neighbors}\\
0 & \text{otherwise}
\end{cases}
\end{equation}
With this definition, and using simple charge continuity considerations, the local electric potential on each island can be written as
\begin{equation}
\boldsymbol{V} = \mathcal{C}^{-1}\left(e\boldsymbol{n}' +\boldsymbol{Q_G}\right)
\label{eq:potential}
\end{equation}
Where $\boldsymbol{V}$ is a column vector with $V_i$ as its entries, and the entries of  $\boldsymbol{n}'$ are defined by
\begin{equation}
en'_i = \begin{cases}
en_i + C_{ix} V_x & \text{island i is a neighbor of electrode x}\\
en_i & \text{otherwise}
\end{cases}
\label{eq:number_of_carriers}
\end{equation}
where $n_i$ is the number of charge carriers on island $i$. $\boldsymbol{Q_G}$ holds the charge on each gate capacitor (with sign convention as indicated in Fig. \ref{fig:2D_model}c).
Using Eq. (\ref{eq:energy_difference}-\ref{eq:number_of_carriers}) and treating the tunneling of charge carriers between neighboring islands as a small perturbation, the tunneling rates are given by \cite{ingold1992}: 

\begin{equation}
\Gamma\left(\Delta E\right) =\frac{1}{e^2R_T} \frac{-\Delta E}{1-\exp\left(\Delta E/k_B T\right)}
\label{eq:rates_low_impedance}
\end{equation}

Where $R_T$ is the tunneling resistance between the islands, and $\Delta E$ is the electrostatic energy difference between the charge configurations before and after the tunneling event. For low temperatures, tunneling will only occur if $\Delta E<0$, namely when the voltage difference between the islands is large enough to compensate for the additional Coulomb energy. Indeed, these previous calculations, which assumed that the charges instantaneously relax to their lowest-energy configuration after the tunneling event, found that this Coulomb blockade leads to a threshold voltage below which current cannot flow through the system \cite{middleton1993}. 

As mentioned above, the assumption of instantaneous relaxation precludes hysteresis, and thus these models cannot explain some of the observations. The importance of a finite relaxation time has been pointed out by Korotkov \cite{korotkov1994}, who showed that for a single island, a finite relaxation time may lead to hysteresis. Here we allow for such a finite relaxation time by connecting each island to a gate or substrate via an RC circuit (dotted square in Fig.~\ref{fig:2D_model}c). This RC coupling is what determines the charge dynamics between tunneling events in our model, and it allows us to easily change the relaxation time. To find the equations describing this process, we calculate the electric potential on each island by its voltage difference from the gate,
\begin{equation}
V_i = V_G - \left[\boldsymbol{R_G}\right]_i\left[\frac{d\boldsymbol{Q_G}}{dt}\right]_i - \frac{1}{\left[\boldsymbol{C_G}\right]_i}\left[\boldsymbol{Q_G}\right]_i
\label{eq:potential_from_gate}
\end{equation}
Solving Eq. (\ref{eq:potential},\ref{eq:potential_from_gate}) for $d\boldsymbol{Q_G}/dt$, we find the following differential equations:
\begin{eqnarray}
&\frac{d\boldsymbol{Q_G}}{dt} =\mathcal{T}^{-1}\left(\boldsymbol{Q_G} - \boldsymbol{Q_n}\right)\nonumber\\
&\left[\mathcal{T}^{-1}\right]_{ij} \equiv \frac{1}{\left[\boldsymbol{R_G}\right]_i} \left(\left[\mathcal{C}^{-1}\right]_{i,j} + \frac{1}{\left[\boldsymbol{C_G}\right]_i}\delta_{i,j}\right)\\
&\left[\boldsymbol{Q_n}\right]_i \equiv \frac{1}{\left[\boldsymbol{R_G}\right]_i}\sum_j \left[\mathcal{T}\right]_{ij}\left(V_G \delta_{ij} + \frac{1}{\left[\boldsymbol{C_G}\right]_j}en_j'\right) - en_i'\nonumber
\label{eq:2D_QG_ODE}
\end{eqnarray}
We can obtain a numerical solution for this linear system of ordinary differential equations (ODEs), via diagonalization of $\mathcal{T}$, and we use it to calculate the dynamics of charge distribution between tunnelings.

\begin{figure*}[ht]
    \includegraphics[width=0.8\textwidth]{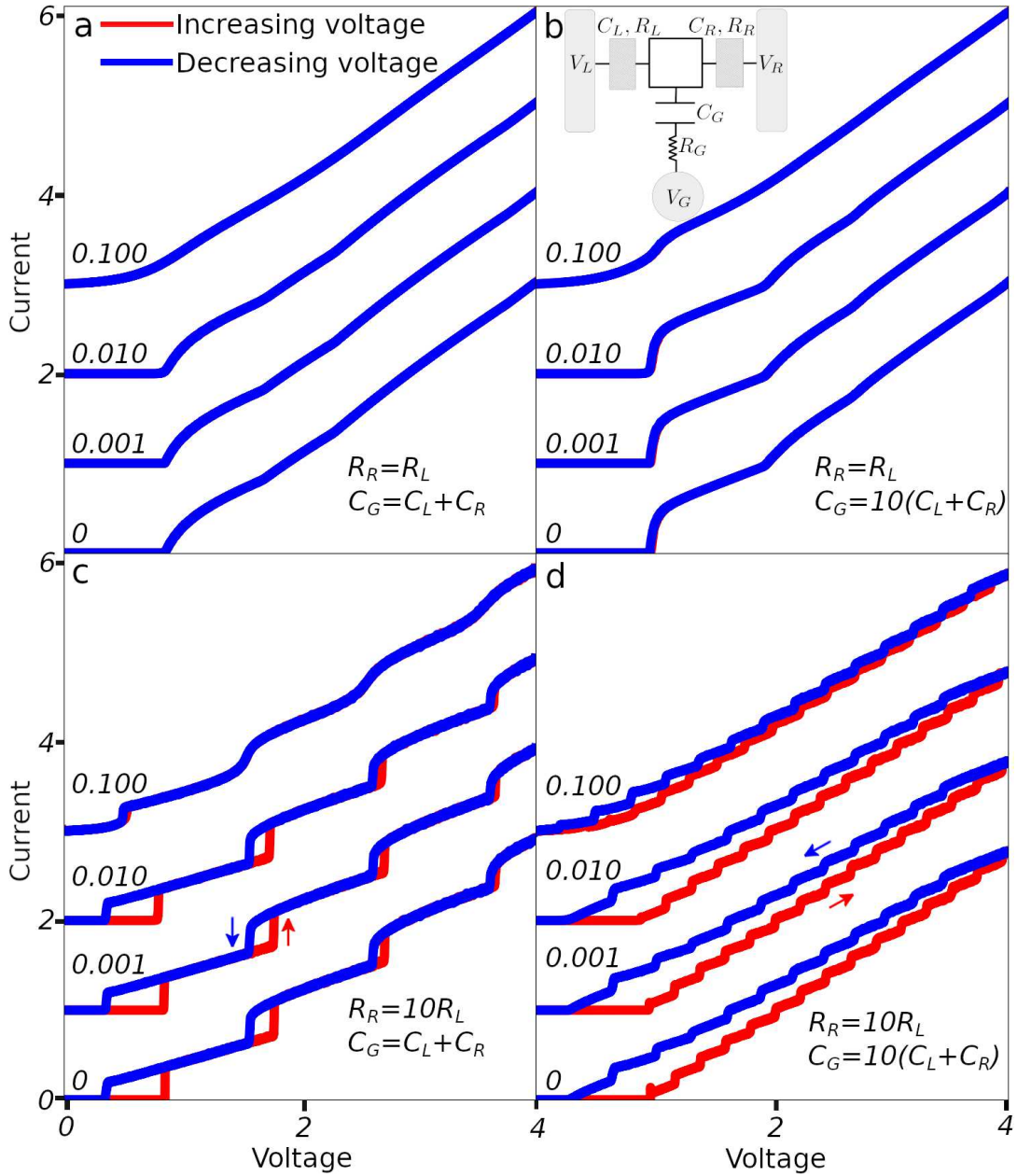}
    \caption[A single island model reproduce hysteresis and jumps for large gate capacitance and tunneling-resistance asymmetry]{
        Simulated single island I-V curves, for different temperatures (temperature is indicated above each curve). Temperature, voltage, and current are expressed in units of $e^2/\left(k_B\left<C\right>\right)$, $e/\left<C\right>$, and $e/\left(\left<R\right>\left<C\right>\right)$, respectively. Curves for different temperatures are offset from one another by $e/\left(\left<R\right>\left<C\right>\right)$.
        (a) Small gate capacitance and equal resistance for both junctions. (b) Large gate capacitance and equal resistance. Inset: Single island model schema.
        (c) Small gate capacitance and asymmetric tunneling resistance.
        (d) Large gate capacitance and asymmetric resistance.
        Simulation parameters: $C_L=2C_R$, $V_G= 0$, $R_G=1000\left(R_L+R_R\right)$.
    }
    \label{fig:single_island_results}
\end{figure*}

For the simulation results presented here, we used the same gate resistance and capacitance for all islands. From Eqs.~(\ref{eq:rates_low_impedance} and \ref{eq:2D_QG_ODE}), the ratio of the relaxation time to the intertunneling time is determined by the ratio of gate resistance $R_G$ to the tunneling resistances $R_T$. Therefore,  in the limit where $R_G$ is much bigger than all relevant tunneling resistances, one cannot employ the instantaneous relaxation assumption used in previous calculations. In the following, we incorporate this finite relaxation time in the simulations.

Assuming that the time between consecutive measurements is much longer than the relaxation time, we can ignore the relaxation dynamics and calculate only the steady-state solution for the current in our model. To do that, we used Kinetic Monte Carlo simulations \cite{gillespie1977}. At each step of our simulation, Eqs.~(\ref{eq:energy_difference}--\ref{eq:rates_low_impedance}) are used to calculate tunneling rates and determine the probabilities for each tunneling event. The next tunneling event, and the time to it, are then chosen accordingly. Charge distribution is updated between tunneling events according to Eq.~(\ref{eq:2D_QG_ODE}). Since tunneling rates depend on the charge distribution (through Eq.~(\ref{eq:potential})) this update is done in small time intervals, updating the chosen next tunneling event, and the time to it, after each interval. To calculate $I-V$ curves, we run the simulation for each value of the external voltage until we reach a steady state (which was verified by checking that the average charge distribution is constant). The steady-state current is then calculated by averaging the relevant tunneling rates in the system. We then proceed to run the simulation for the next voltage value, starting from the charge distribution as it was at the end of the run for the previous voltage value.

\section{Results \label{results}}
\subsection{A Single Island}

In order to understand the role of the different parameters, let us first study a simple system, consisting of a single conducting island (see inset in Fig. \ref{fig:single_island_results}b), for which one can also obtain analytical results for $T=0$ (for detailed analytical solution, see Appendix \ref{single_island_analytic}).  These $T=0$ exact results and our simulation results, for $T>0$, are presented in Fig. \ref{fig:single_island_results}. Comparing results for an island with and without  asymmetry between the left and right tunneling resistances (Fig. \ref{fig:single_island_results}(a) and \ref{fig:single_island_results}(b) versus Figs. \ref{fig:single_island_results}(c) and \ref{fig:single_island_results}(d)), we conclude that, in order to get  visible hysteresis and current jumps, tunneling resistance asymmetry ($R_L \neq R_R$) is essential. In addition, for the cases with hysteresis, we see that hysteresis loop area becomes bigger when $C_G$ increases (Fig. \ref{fig:single_island_results}(c) versus \ref{fig:single_island_results}(d)). As temperature or applied voltage increase, the area of each hysteresis loop becomes smaller, and the $I-V$ curves become smoother.  It is encouraging to note that these results qualitatively resemble the experimental $I-V$ curves for a single quantum dot \cite{duruoz1995}.
We can understand those results intuitively in the following way: Tunneling resistance asymmetry will control the average number of charge carriers in the system. For example, if $R_R > R_L$ it would be ``easier" for carriers to tunnel into the island from the left electrode than it is to tunnel from the island to the right electrode, and the average number of carriers on the island would grow with the external voltage. This would, in turn, change $Q_G$, which is the slow degree of freedom, which acts as a memory in our system. Increasing $C_G$, in addition, would enable more steady-state solutions for $Q_G$  and thus more jumps and larger hysteresis loops (see Appendix \ref{single_island_analytic}).

\begin{figure*}[ht]
    \includegraphics[width=0.85\textwidth]{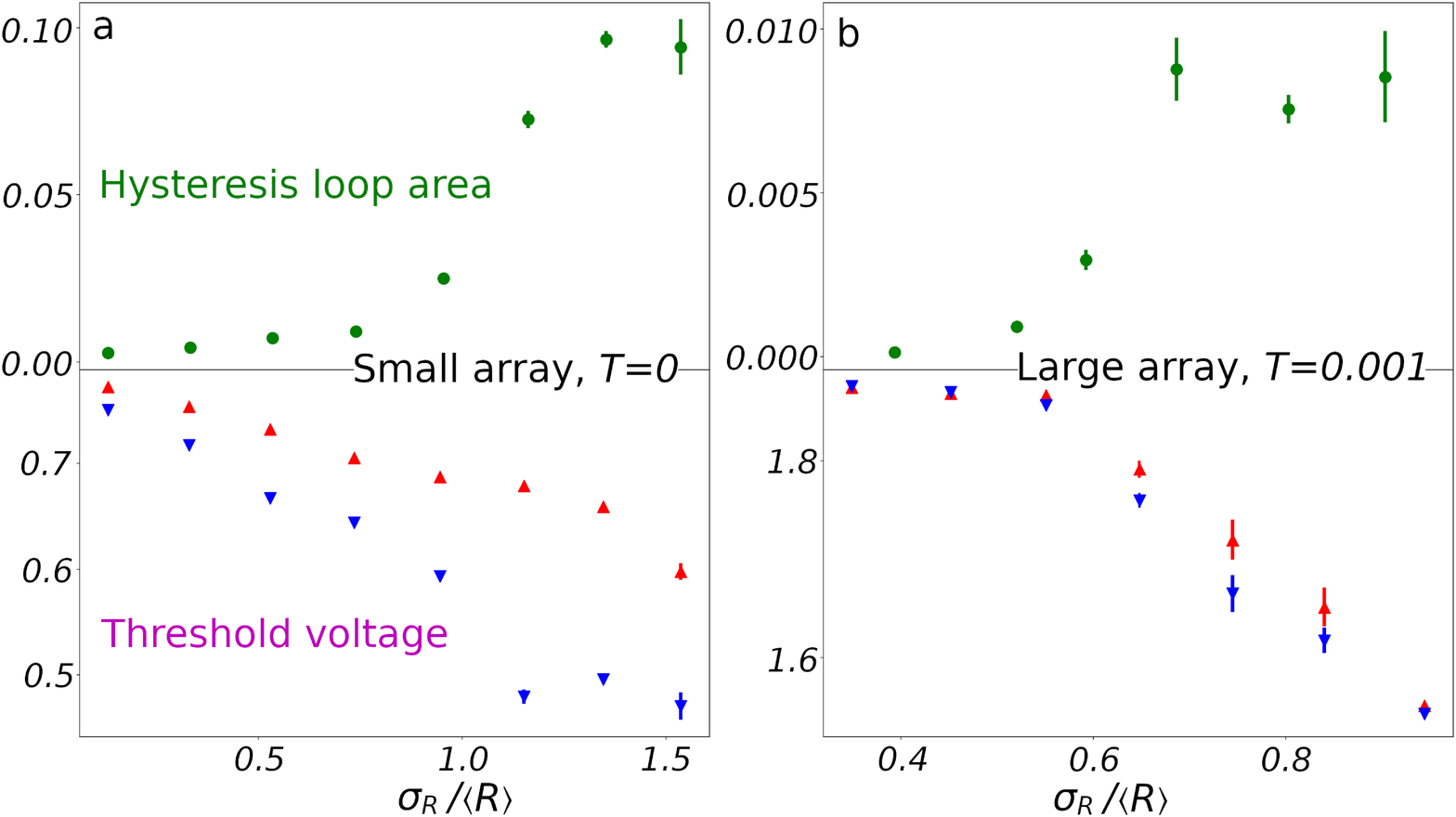}
    \caption[When resistance disorder increases, hysteresis loop area increases and threshold voltages decrease]{
        Hysteresis loop area and threshold voltages as a function of resistance disorder. Calculated by averaging the results from $>10$ different array realizations, with various tunneling capacitance disorder values, for each data point. Error bars show standard error of averages. Upward-pointing red triangles, increasing voltage. Downward-pointing blue triangles, decreasing voltage. Temperature, loops areas, and threshold voltages are expressed in units of $e^2/\left(k_B\left<C\right>\right)$, $e^2/\left(\left<C\right>^2\left<R\right>\right)$, and $e/\left<C\right>$, respectively.
        (a) For $3\times3$ arrays at $T=0$.
        (b) For $10\times10$ arrays at  $k_B T=10^{-3}\times e^2/\left<C\right>^2$.
        Common simulation parameters:   $V_G= 0$, $R_G=100\left<R\right>$, $C_G = 10\left<C\right>$.
    }
    \label{fig:different_disorder_results}
\end{figure*}

\subsection{Normal Array}
Following the single island results, we now turn to calculating $I-V$ curves for normal disordered arrays with large gate capacitance ($C_G > \left<C\right>$). Disorder was realized by choosing, for each junction, random tunneling resistance and capacitance. The choice of tunneling resistance distribution is motivated by the SC island geometry. As the  tunneling resistance between islands depends exponentially on the  distance between them, which we expect to be smoothly distributed,  we set the tunneling resistances  to be $e^\alpha$, where $\alpha$ is a uniform random variable (which is effectively proportional to the distance between islands). 

For a single island, we saw that tunneling resistance asymmetry had a significant effect on hysteresis area. Therefore, it is natural to expect that tunneling resistance disorder would play a major role in determining the hysteresis loop area in arrays. To check this, we ran our simulations for different disorder realizations. We then grouped together realizations with a similar standard deviation for the tunneling resistance distribution (at least $10$ realizations at each group). For each group, we calculated the average hysteresis area and the threshold voltages (both for increasing and decreasing applied voltage). The results are plotted in Fig.~\ref{fig:different_disorder_results}. Indeed, as could be expected from the single island results, resistance disorder has a crucial effect on hysteresis area. For small disorder, hysteresis almost vanishes as we get a very small loop area. When disorder increases, the hysteresis loop area increases as well. Interestingly, both threshold voltages drop when disorder increases (despite being resistance-independent for a single island). As disorder increases, the difference between the threshold voltages for increasing and decreasing voltages tend to increase, matching the increase in hysteresis area. This is more apparent for the small arrays (Fig. \ref{fig:different_disorder_results}(a)) than it is for the larger ones (Fig. \ref{fig:different_disorder_results}(b)).

Next we checked the effect of temperature. In Fig. \ref{fig:2D_model}(d), we compare I-V curves, of the same $10\times 10$ array realization, for different temperatures. As temperature increases, the threshold voltage and hysteresis loop area become smaller. The big hysteresis loop we see for small temperatures diminishes when temperature rises, and we are left with smaller loops, which also disappear for even higher temperatures. This behavior is in qualitative agreement with the experimental results for both the QD array \cite{duruoz1995}, Fig.~\ref{fig:2D_model}(a)   and the disordered SC film \cite{cohen2011}, Fig. \ref{fig:2D_model}(b).  Moreover, in agreement with the low-temperature experimental results, the threshold voltage for increasing voltage is temperature-dependent, while that for decreasing voltage is not. 
For a more detailed comparison with the results obtained for the QD array \cite{duruoz1995} (Fig.~\ref{fig:2D_model}(a)), one can notice that each hysteresis loop is composed of a single abrupt current jump for increasing voltage and a few smaller jumps when decreasing it. This is in agreement with our simulation results (Fig. \ref{fig:2D_model}(d)) and in contrast to the results for SC films  \cite{cohen2011} (Fig.~\ref{fig:2D_model}(b)), where we do see multiple jumps in the increasing branch of the hysteresis loop for some temperatures ($75 < T < 130 mK$). In the low temperature QD-array results \cite{duruoz1995} (Fig. \ref{fig:2D_model}(a)), we do see a second, smaller, hysteresis loop, for small voltages, which is not apparent in our simulation results. This could be the result of array size differences ($200 \times 200$ array in the experiment versus $10 \times 10$ in our simulation) or it can be an outcome of the specific disorder realization.

\begin{figure*}[ht]
    \includegraphics[width=0.85\textwidth]{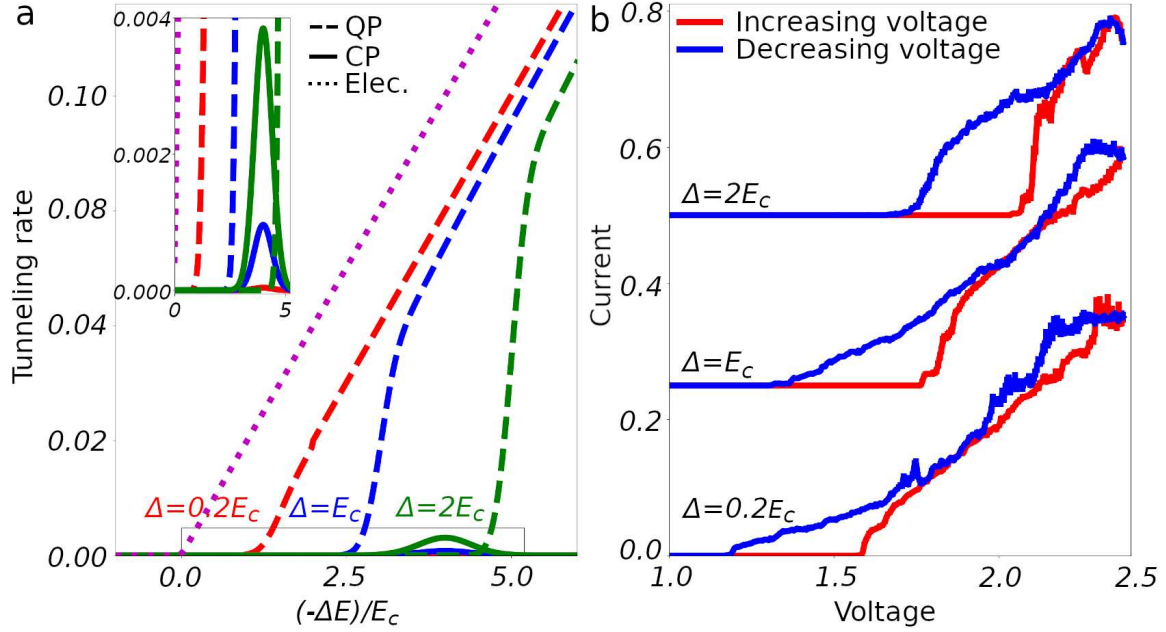}
    \caption[Superconducting arrays show bigger jumps and hysteresis for a larger SC gap]{
    (a) Tunneling rates for Cooper pairs (solid lines) and quasiparticles (dashed lines) for different superconducting gaps. The tunneling rates for electrons (dotted line) are plotted for comparison.
    Inset: Closeup view of the marked region. Tunneling rates are expressed in units of $1/\left(\left<R\right>\left<C\right>\right)$.
    (b) $I-V$ curves for a $5 \times 5$ superconducting array with different superconducting gaps. Error bars indicate standard errors. Voltage and current are expressed in units of $e/\left<C\right>$ and $e/\left(\left<R\right>\left<C\right>\right)$, respectively. Curves are shifted up by $0.25e/\left(\left<R\right>\left<C\right>\right)$ from each other.
    Simulation parameters: $k_B T=0.4\cdot 10^{-3}e^2/\left<C\right>$, $E_c = 25\cdot 10^{-3}e^2/\left<C\right>$, $V_{max}=2.4\left<C\right>/e$, $V_G= 0$, $R_G=234\left<R\right>$, $C_G = 21\left<C\right>$, $\sigma_C = 0.64\left<C\right>$, $\sigma_R = 0.78\left<R\right>$.
    }
    \label{fig:sc_results}
\end{figure*}

\subsection{Superconducting Array}

Having demonstrated that by incorporating a finite relaxation time in calculations of the transport through a random normal array we can explain the experimental observations of threshold, multiple jumps and hysteresis, we next checked how these results are modified when considering transport through a superconducting array. In this case, both quasi-particles and Cooper-pairs can, in principle, tunnel between the islands, and one has to take into account their different statistics and density of states. The resulting tunneling rates are \cite{ingold1992}
\begin{equation}
\Gamma_{cp}\left(\Delta E\right) = \frac{\pi}{2\hbar} E_J^2 P_2\left(-\Delta E\right)
\label{eq:cooper-pair_tunneling_rate},
\end{equation}
for Cooper-pairs, where $E_J$ is Josephson's energy, obtained from $\Delta$ using the standard relation \cite{tinkham_introduction_2004}, and  $P_\kappa\left(E\right)$ is the probability of absorbing energy $E$ by the environment, $\kappa=1$ for quasi-particles, and $\kappa=2$ for Cooper-pairs. For quasi-particles
\begin{eqnarray}
\Gamma_{qp}\left(\Delta E\right) =& \frac{1}{e^2 R}\int_{-\infty}^{\infty} dE \int_{-\infty}^{\infty} dE' \frac{N_s\left(E\right) N_s\left(E'-\Delta E\right)}{N\left(0\right)^2} \times\nonumber \\
&f\left(E\right)\left[1-f\left(E'-\Delta E\right)\right]P_1\left(E-E'\right) ,
\label{eq:quasi-particle_tunneling_rate}
\end{eqnarray}
where $f\left(E\right)$ is Fermi-Dirac distribution. 
$N_s\left(E\right)/N\left(0\right)$ is the  BCS quasiparticle density of states, which is given by \cite{bardeen1957}
\begin{equation}
\frac{N_s\left(E\right)}{N\left(0\right)} = \begin{cases} \frac{\left|E\right|}{\sqrt{E^2-\Delta^2}} & \left|E\right|>\Delta\\
0 & \left|E\right| <\Delta \end{cases}
\label{eq:quasi_particle_dos}
\end{equation}

For a high impedance environment, where the perturbation theory leading to Eq.~\ref{eq:cooper-pair_tunneling_rate} is valid, $P_{\kappa}\left(E\right) $ is given by  \cite{ingold1992}
\begin{equation}
P_{\kappa}\left(E\right) = \frac{\exp\left[-\left(E-\kappa^2E_c\right)^2/4\kappa^2E_ck_BT\right]}{\sqrt{4\pi \kappa^2E_c k_B T}}
\label{eq:energy_absorption_high_impedance}
\end{equation}
where $E_c$ is the charging energy of the environment.

Tunneling rates were calculated, in a similar way to \cite{cole2014}, assuming that energy is absorbed by the intrinsic impedance of the array. In that case, the dominant impedance comes from the gate coupling ($R_G$ in the slow relaxation limit), and the environment's charging energy ($E_c$ in Eq.~\ref{eq:energy_absorption_high_impedance}) is given by $e^2/2C_G$. The calculated tunneling rates are plotted in Fig. \ref{fig:sc_results}(a). Cooper-pair tunneling rates (solid curves) are centered around $\left(-\Delta E\right)\approx 4E_c$ which is the energy that is best absorbed by the intrinsic impedance (see eq. \ref{eq:energy_absorption_high_impedance}). Quasiparticles require $\left(-\Delta E\right) > \max\left[2\Delta, E_c\right]$ to overcome the SC gap and, in addition, to compensate for the energy that is absorbed by the intrinsic impedance (dashed curves). For larger energy gains, rates grow linearly, in a similar way to the tunneling rates of electrons in normal arrays (purple dotted curve). For $\Delta \approx 2E_c$, both cooper pairs and quasiparticles tunneling would give a significant contribution to the current. For $\Delta \ll E_c$, only quasi-particles would contribute to the current.

We incorporated these tunneling rates into our simulation to calculate the current through SC arrays. The resulted $I-V$ curves, for a $5\times 5$ array, are plotted in Fig. \ref{fig:sc_results}(b). For all superconducting gap values, we see hysteresis and jumps. Increasing $\Delta$  we get a larger threshold voltage, and we see a bigger current jump when the external voltage is increased above it. Unlike the results for a normal array (Fig. \ref{fig:2D_model}(d)), the hysteresis loops obtained for SC array are composed of multiples current jumps, both for increasing and decreasing applied voltage. Those jumps are bigger for the increasing branch of the hysteresis loop, and smaller for the decreasing branch. This is in qualitative agreement with the experimental results for SC films \cite{cohen2011} (Fig. \ref{fig:2D_model}(b)).

\section{Discussion and conclusion\label{conclusion}}

The validity of our model for SC films requires the existence of well-defined islands, where Cooper pairs and quasiparticles can be localized. For a SC state to be generated in a single island, the average energy spacing in it must be smaller than the SC gap  \cite{gantmakher2010}
\begin{equation}
\delta \epsilon = \left(N\left(0\right)b^2\right)^{-1} < \Delta
\label{eq:SC_islands_condition}
\end{equation}
where $b$ is the lengthscale for an island, and $N\left(0\right)$ is the two-dimensional density of states on the Fermi level. In addition, the tunneling rates calculations for Cooper pairs involve the Josephson coupling term in the Hamiltonian \cite{ingold1992}, which assumes a defined phase for the SC wave function on each island (for a detailed explanation of this point, see \cite{efetov1980}). This requires the assumption that the island length is smaller than the coherence length $\xi$. These requirements impose the following bounds for an island size:
\begin{equation}
\left(N\left(0\right)\Delta\right)^{-1/2} < b < \xi
\label{eq:SC_islands_size_bound}
\end{equation}
Our analysis further requires that the tunneling term can be treated as a perturbation. For the normal case (or for quasiparticles), this requirement is satisfied if tunneling resistances are bigger than the resistance quantum, $\hbar/e^2$ \cite{ingold1992}. For Cooper-pair tunneling, since the tunneling term in the Hamiltonian is proportional to the Josephson energy, $E_J$, the requirement is $E_J \ll E_c$. This assumption should be satisfied for the insulating side of the SIT, but it might break when approaching the phase transition. In addition, in our simulation of the slow relaxation limit we assume a timescale hierarchy that is, from longest to shortest, as follows: measurement time, relaxation time, time between tunneling events, and time for the tunneling itself. 
For a crude estimation of the required relaxation time to satisfy our model's assumptions, we used the measured current in \cite{cohen2011,duruoz1995} to estimate an average intertunneling time of $\sim 10^{-10}-10^{-12}$ s. This means that the relaxation time should be at least an order of magnitude longer, at least a nanosecond, and the measuring frequency should be much slower than a $GHz$ to always measure a steady state. The relaxation timescale could be determined by the coupling (e.g resistance and capacitance) between the sample and the substrate or gates. Since this is usually uncontrolled in experiments, it may explain why some samples show hysteresis and jumps, while others, nominally with the same parameters, do not (see, e.g. \cite{doron2020}). Another possible source of slow dynamics may be two-level systems,  which have already been observed \cite{wissberg2018,kremen2018} in similar disordered SC films. 

It is interesting to note that similar results have also been obtained for SC films and for a JJA in the superconducting side of the SIT \cite{bottcher2018, levinson2019, doron2020}, but with $I$ and $V$ interchanged (i.e., observing voltage jumps and hysteresis when modifying the external current). This is more speculative, but it is possible that our model is also relevant for the SC side of the transition, relying on the duality between charge and phase in superconductors, as suggested and discussed in \cite{vinokur2008, baturina2013}.

It is important to clarify that our model is not contradictory to the electron-overheating model. We merely show that the observed $I-V$ curve features could be accounted by a random array model, which is in thermal equilibrium. It is possible that a more accurate model could be achieved by combining the two approaches together, which would require some further work. 

Our results point out the importance of different timescales in such disordered strongly interacting systems, and offer a novel viewpoint: the system might not be in thermal and electrostatic equilibrium but rather in a new type of steady state, depicted by the slow degree of freedom in the system. It would be interesting, though experimentally challenging, to control the relaxation rate and check the predictions of our theory, showing, for example, the suppression of the hysteretic behavior for a short relaxation time.

To conclude, we have introduced an ingredient into the physics of disordered systems --  a slow relaxation of some of the degrees of freedom, in this case the charge distribution, towards electrostatic equilibrium. It has been demonstrated that including this ingredient into a random array model explains the puzzling observations of hysteresis and multiple current jumps in $I-V$ curves, as measured in disordered superconductors near the superconductor-insulator transition, and in quantum-dot arrays. Our results support the SC islands formation scenario as the correct description for the SIT in amorphous SC films, and adding an important tier to our understanding of disordered quantum systems in general. 

Our model, and simulation method, can be used as a basis for future research on disordered materials where relaxation cannot be considered instantaneous. In addition, the approach we used, i.e., adding a slow degree of freedom to account for hysteresis in a system, might be applicable to other, possibly hysteretic, disordered systems \protect\footnote{Our simulation is written in Python and is available as an open-source code at \url{https://github.com/kasirershaharbgu/random_2D_tunneling_arrays}}.

\begin{acknowledgments}
The authors acknowledge useful discussions with A. Erez, D. Shahar, B. Weiner,  N.~S. Wingreen  and L. Zhu. 
This work was supported by the Israel Science Foundation (grant 3523/2020).
\end{acknowledgments}

\renewcommand{\theequation}{A.\arabic{equation}}

\newpage
\appendix
\section{Single island at zero temperature - analytic solution \label{single_island_analytic}}

\setcounter{equation}{0}

\begin{figure}[ht]
    \includegraphics[width=\columnwidth]{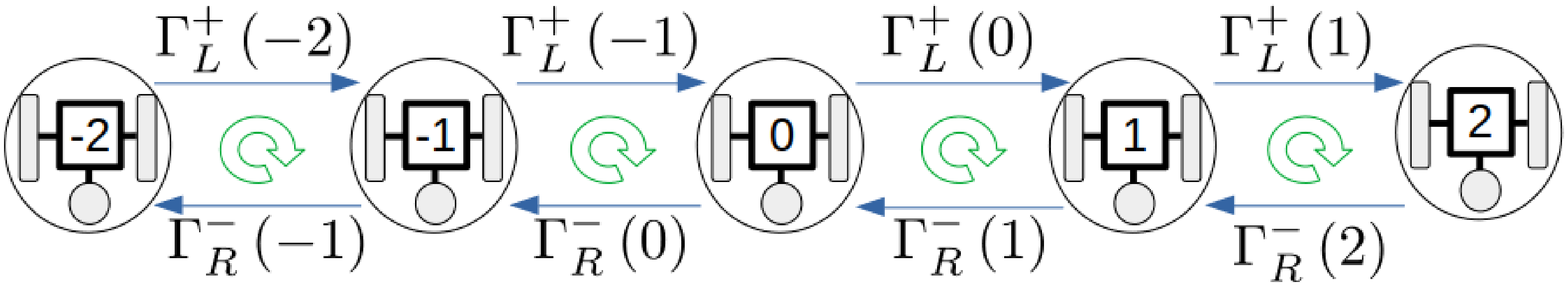}
    \caption[Single island graph representation]{Single island possible states graph representation. \small Calculated for $C_L=C_R$,$V_G=0$ and small external voltage. Each vertex represents an occupation state. The number of excess carriers on the island is indicated for each state. Edges connect neighboring states and their weight corresponds to the matching tunneling rate. $\Gamma^+_x$,$\Gamma^-_x$ are the rates for adding or removing one electron onto(from) the island from(onto) electrode $x$,  respectively.  Green arrows show cycles for which one electron charge passes from left to right.}
    \label{fig:states_graph_single_island}
\end{figure}

In the case of a single island at zero temperature (see the schematic in Fig. \ref{fig:single_island_results}), we can represent the possible states and transitions of our system in a simple graph in which each state is represented by a vertex, and the weight of each directed edge is the matching transition rate (e.g. Fig. \ref{fig:states_graph_single_island}). We would like to calculate the probability of finding the system in each state. To do so, we define the outflow from each vertex to be its probability times the sum of all its outgoing edges' weights. We define the inflow to a vertex as the sum of outflows from all vertices which have an edge from them to that vertex. In a steady state, the inflow and outflow of each vertex are equal.  For a given $Q_G$ and $V$, one can find the steady-state solution using the following procedure: First, an initial probability is assigned to an edge vertex (e.g. $n=\pm2$ in Fig. \ref{fig:states_graph_single_island}). Then, other probabilities are calculated such that the inflow and outflow are equal for each vertex, starting from the chosen vertex and gradually going to the opposite side of the graph. Finally, the probabilities are normalized such that their sum would be $1$. Having calculated the steady-state probabilities, we can calculate the average current by
\begin{eqnarray}
I\left(Q_G,V\right) = -e\sum_n p\left(n\right) \left[\Gamma_L^+\left(n\right) - \Gamma_L^-\left(n\right)\right] \nonumber\\= -e\sum_n p\left(n\right) \left[\Gamma_R^-\left(n\right) - \Gamma_R^+\left(n\right)\right]
\label{eq:steady_state_current_single_island}
\end{eqnarray}
Notice that the steady-state probability and the tunneling rates depend both on $Q_G$ and $V$.

\begin{figure}[h]
    \includegraphics[width=0.85\columnwidth]{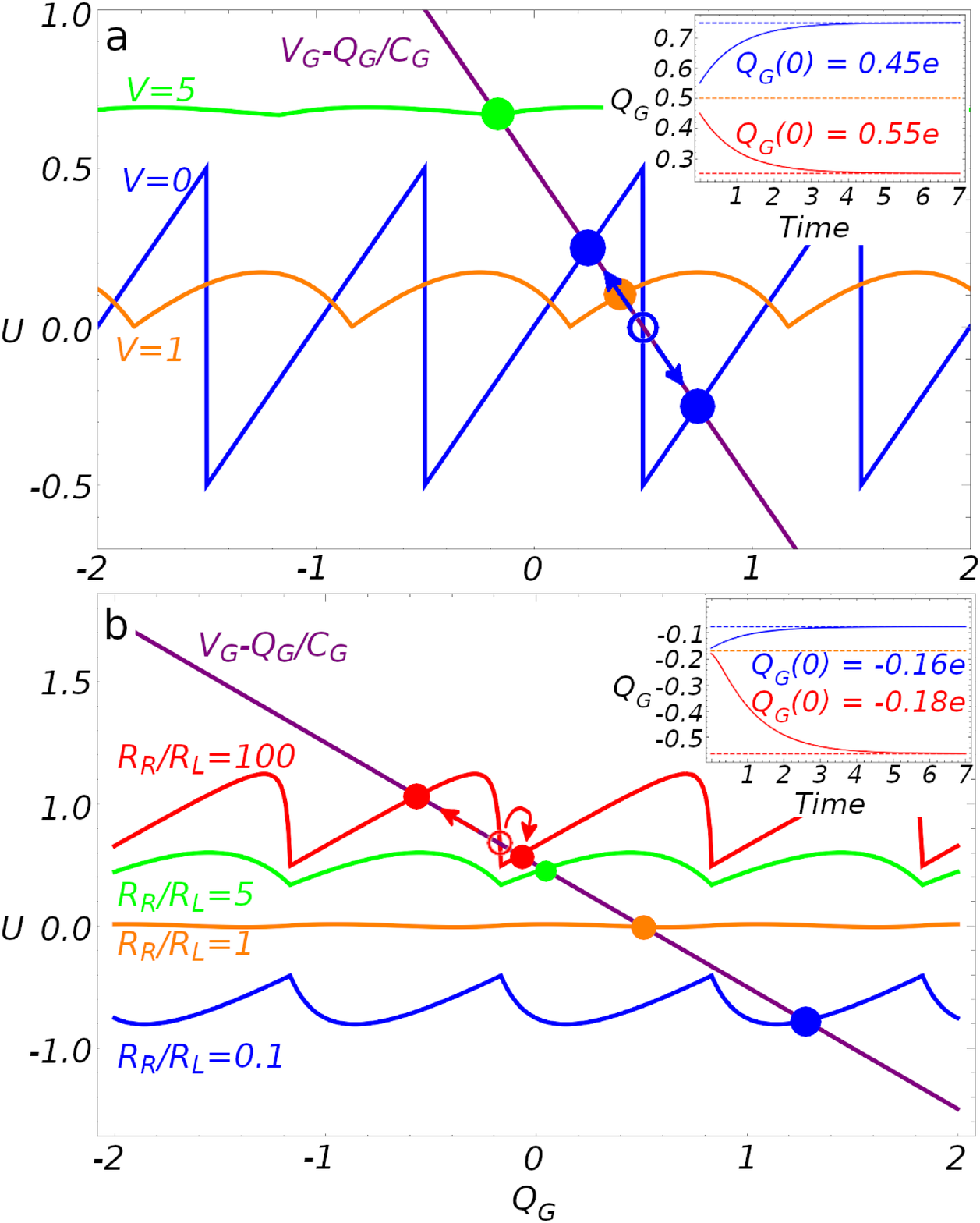}
    \caption[Average voltage on island]{Average voltage on island \small for different external voltages (a) and resistances (b). Stable steady-state solutions are marked with a whole circle, unstable with an empty circle.  Small arrows indicate the direction of convergence for bistable cases ($V=0$ in (a) and $R_R/R_L=100$ in (b)). Voltages are expressed in units of $\frac{e}{C_L+C_R}$.
    (a) For $V=0$, the shape is sawtooth-like. For large $V$, the average $U$ approaches a $Q_G$ independent solution.
    (b) As the resistances ratio, $R_R/R_L$, increases, $\left<U\right>_{Q_G}$ shape resembles the $V=0$ result. For $R_L > R_R$ the result is mirrored about the $x$-axis.
    Insets: $Q_G$ for bistable cases, as a function of time, for different initial values. Dashed lines mark the steady-state solutions (red and blue,stable; orange, unstable).
    Mutual parameters: $C_R = 2C_L$, $C_G = C_L+C_R$, $V_L = -V_R$, $V_G = 0.5\frac{e}{C_L+C_R}$. a $R_R=2R_L$. b $V=2\frac{e}{C_L+C_R}$.}
    \label{fig:graphical_sol}
\end{figure}

The resulting current depends on $Q_G$. For slow relaxation, $Q_G$ will relax to its mean equilibrium value. In that case, the solution for Eq. \ref{eq:2D_QG_ODE} would be 

\begin{eqnarray}
Q_G\left(n,t\right) =& \left(Q_0-\left<Q_n\right>_{Q_G}\right)e^{-\frac{t}{\tau}} + \left<Q_n\right>_{Q_G}\nonumber\\
\left<Q_n\right>_{Q_G} \equiv& \sum_n p_{Q_G}\left(n,V\right) Q_n\left(V\right)
\label{eq:QG_sol_slow_relaxation}
\end{eqnarray}
The $Q_G$ subscript is there to remind us that the steady-state probability, and hence also all averages, depend on $Q_G$.
The steady state solution for $Q_G$ would satisfy
\begin{equation}
Q_G = \left<Q_n\right>_{Q_G}
\label{eq:steady_state_QG_slow_relaxation}
\end{equation}
This equation can have multiple solutions for a given $V$. In that case, it is possible to get hysteresis in the $I-V$ curve. To understand the conditions for hysteresis, we write Eq. \ref{eq:steady_state_QG_slow_relaxation} in terms of $U$, the electric potential on our single island. Using Eq. \ref{eq:potential_from_gate} and \ref{eq:2D_QG_ODE}, we get
\begin{equation}
\left<U\right>_{Q_G} = V_G - \frac{Q_G}{C_G}
\label{eq:steady_state_potential_slow_relaxation}
\end{equation}
This is the same as requiring that, for a steady state, the average current through $R_G$ would vanish.

Both sides of Eq. \ref{eq:steady_state_potential_slow_relaxation} are plotted in Fig. \ref{fig:graphical_sol}. Hysteresis is possible if this equation has multiple solutions. This is more likely for a large $C_G$ when the linear function on the right-hand side of Eq. \ref{eq:steady_state_potential_slow_relaxation} has a smaller slope. The exact condition is
\begin{equation}
C_G > \left(\max_{Q_G}\left[-\frac{\partial \left<U\right>_{Q_G}}{\partial Q_G}\right]\right)^{-1}
\label{eq:hysteresis_condition_for_potential}
\end{equation}
since otherwise the linear function $V_G-\frac{Q_G}{C_G}$ (purple curves in fig. \ref{fig:graphical_sol}) decreases faster than $\left<U\right>_{Q_G}$  anywhere, and thus crosses it only once (e.g., green curves in fig. \ref{fig:graphical_sol}). Analysis  of $\left<U\right>_{Q_G}$ for small voltages, such that only 2 states are possible, shows that we will get hysteresis for
\begin{equation}
C_G > \left(C_L + C_R\right)\left[\frac{1}{1-\frac{R_{min}}{R_{max}}\frac{\left(C_L+C_R\right)V}{e}}-1\right]
\label{eq:hysteresis_cond_small_v_main_text}
\end{equation}
We can rewrite this, to get a condition for $V$:
\begin{equation}
V\frac{C_L+C_R}{e} < V^{hyst}\frac{C_L+C_R}{e} \equiv \frac{R_{max}}{R_{min}}\frac{C_G}{C_\Sigma}
\label{eq:hysteresis_condfor_v_small_v_main_text}
\end{equation}

As seen in Fig. \ref{fig:graphical_sol}(a), for a small voltage (blue curve), we obtain more than one steady-state solution. For a larger voltage (orange curve) there is only one solution. For an even larger voltage (green curve), more states become available (more carriers can tunnel to or from the island, and $\left<U\right>_{Q_G}$ becomes approximately $Q_G$-independent). This convergence is slower when the ratio $R_{\max}/R_{\min}$ is bigger, as shown in fig. \ref{fig:graphical_sol}(b). For $R_L=R_R$ (orange curve), $\left<U\right>_{Q_G}$ is almost $Q_G$-independent for $V = 2
\frac{e}{C_L+C_R}$. For $R_R > R_L$, and the same voltage (green and red curves), $\left<U\right>_{Q_G}$ becomes more similar to the result for low voltages (blue curve in fig. \ref{fig:graphical_sol}(a)). Therefore, we will see a more pronounced hysteresis for small voltages and large resistance ratios.

\renewcommand{\theequation}{B.\arabic{equation}}
\setcounter{equation}{0}

\section{$I-V$ results in a semi-log scale \label{log_scale_results}}

\begin{figure}[h]
    \includegraphics[width=\columnwidth]{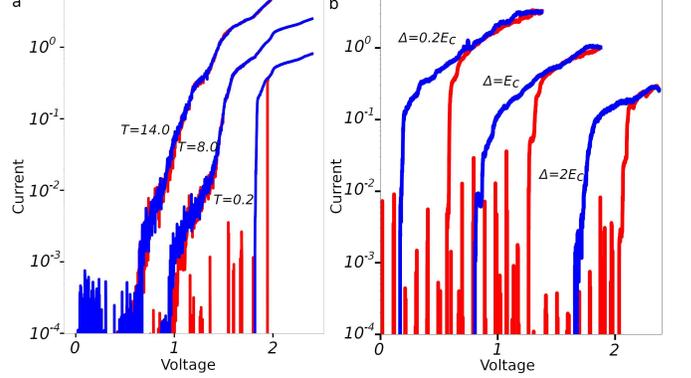}
    \caption[Log-scale results]{Simulated $I-V$ results in a semilog scale. Results become noisy for small current values as a result of the finite numerical accuracy. (a) $10\times 10$ normal array at different temperatures. Selected examples from Fig. \ref{fig:2D_model}(d).  Temperature, voltage, and current are expressed in units of $e^2/\left(k_B\left<C\right>\right)\cdot 10^{-3}$, $e/\left<C\right>$, and $e/\left(\left<R\right>\left<C\right>\right)$, respectively. $\left<C\right>$ and $\left<R\right>$ are the average tunneling junctions capacitance and resistance, respectively. Curves are offset from one another via multiplication by 
    $3$.
    Simulation parameters:  $V_{max}=2.4\left<C\right>/e$, $V_G= 0$, $R_G=50\left<R\right>$, $C_G = 5\left<C\right>$, $\sigma_C = 0.05\left<C\right>$, $\sigma_R = 0.9\left<R\right>$.
    (b) $5 \times 5$ superconducting array for different superconducting gap values; same results as in fig. \ref{fig:sc_results}(b). Voltage and current are expressed in units of $e/\left<C\right>$ and $e/\left(\left<R\right>\left<C\right>\right)$, respectively. Curves are shifted left by $e/\left<C\right>$ and up via multiplication by $3$ from each other.
    Simulation parameters: $k_B T=0.4\cdot 10^{-3}e^2/\left<C\right>$, $E_c = 25\cdot 10^{-3}e^2/\left<C\right>$, $V_{max}=2.4\left<C\right>/e$, $V_G= 0$, $R_G=234\left<R\right>$, $C_G = 21\left<C\right>$, $\sigma_C = 0.64\left<C\right>$, $\sigma_R = 0.78\left<R\right>$.}
    \label{fig:log_scale_results}
\end{figure}

Many of the experimental $I-V$ curves in the literature are plotted on a semilog scale. We add here selected simulation results, plotted on a semilog scale (Fig. \ref{fig:log_scale_results}), for a more convenient comparison.  

\newpage
\bibliographystyle{apsrev4-2}
\bibliography{prb_article.bib}

\end{document}